\documentclass[useAMS,usenatbib]{mn2e}

\usepackage{amsmath,amssymb,amsbsy,bm}

\usepackage{graphicx}
\usepackage[usenames]{color}

\usepackage[hypertex]{hyperref}

\definecolor{myred}{rgb}{0.85,0.08,0}
\definecolor{mydb}{rgb}{0,0.08,0.8}

\newcommand{\beq}{\begin{equation}}
\newcommand{\eeq}{\end{equation}}
\newcommand{\bea}{\begin{eqnarray}}
\newcommand{\eea}{\end{eqnarray}}
\def\bi{\begin{itemize}}
\def\ei{\end{itemize}}

\title[CMB TE and PGWs II: Wiener Filtering and Tests Based on Monte Carlo Simulations]{CMB Temperature Polarization Correlation and 
Primordial Gravitational Waves II: Wiener Filtering and Tests Based on 
Monte Carlo Simulations}
\author[N.J. Miller, B.G. Keating, and A.G. Polnarev]{N.J. Miller$^{1}$\thanks{E-mail:nmiller@physics.ucsd.edu}, 
B.G. Keating$^{1}$, and A.G.Polnarev$^{2}$ \\
$^{1}$University of California, San Diego, Mail Code 0424, La Jolla, CA, 92093-0424, United States of America \\
$^{2}$Queen Mary, University of London, Mile End Road, London, United Kingdom}

\pagerange{\pageref{firstpage}--\pageref{lastpage}} \pubyear{2007}

\def\LaTeX{L\kern-.36em\raise.3ex\hbox{a}\kern-.15em
    T\kern-.1667em\lower.7ex\hbox{E}\kern-.125emX}

\begin{document}

\label{firstpage}

\maketitle

\begin{abstract}
In this paper we continue our study of CMB TE cross correlation as a source of information about primordial gravitational waves. In an 
accompanying paper, we 
considered the zero multipole method. In this paper we use Wiener filtering of the CMB TE data to remove the density perturbation 
contribution to the TE power spectrum. In principle this leaves only the contribution of PGWs. We examine 
two toy experiments (one ideal and one more realistic), to see how well they constrain PGWs using the TE power spectrum. We consider 
three tests applied to a combination of 
observational data and data sets generated by Monte Carlo simulations: (1) Signal-to-Noise test, (2) sign test, and (3) Wilcoxon rank sum test. We
compare these tests with each other and with 
the zero multipole method. Finally, we compare the signal-to-noise ratio of TE correlation measurements first with corresponding signal-to-noise 
ratios for BB ground based measurements and later with 
current and future TE correlation space measurements. We found that an ideal TE correlation experiment 
limited only by cosmic variance can detect PGWs with a tensor-to-scalar ratio $r=0.3$ at $98\%$ confidence 
level with the $S/N$ test, $93\%$ 
confidence level with the sign test, and $80\%$ confidence level for the Wilcoxon rank sum test. We also compare all results with corresponding 
results obtained using the zero multipole method. We demonstrate that to measure PGWs by their contribution to the TE cross correlation power 
spectrum in a realistic ground based experiment when real instrumental noise is taken into account, the tensor-to-scalar ratio, $r$, must be 
approximately four times larger. In the sense to detect PGWs, the zero multipole
method is the best, next best is the S/N test, then the sign test, and the
worst is the Wilcoxon rank sum test.

\end{abstract}


%
%

\section{Introduction}

Primordial gravitational waves (PGWs) (tensor) generate negative temperature-polarization (TE) correlation for low multipoles, 
while primordial density (scalar) 
perturbations generate positive TE correlation for low multipoles (see \cite{crittenden94,baskaran06,grishchuk07,pmkpaper1}; and references in 
\cite{pmkpaper1}). 
This signature can be to detect PGWs. The test based on this signature (the \textit{zero multipole method}, see \cite{pmkpaper1}) is useful 
as an insurance against false detection or as a monitor of imperfectly subtracted systematic effects.

In this paper, we analyze an alternative method. This method uses Wiener filtering to remove the 
contribution of density perturbations to the TE cross correlation power spectrum for small $\ell$, leaving only the negative residual component 
of the TE power spectrum due to PGWs. 
Actually, this method can be treated a test of the (negative) contribution to the TE correlation power spectrum due to PGWs on large scales 
using uncertainties in the measurements 
consistent with the total TE power spectrum. We use Monte Carlo simulations to analyze the 
probability of detecting PGWs using this method.

By detection of PGWs we mean in this paper, the measurement of the parameter $r$, the ratio of the primordial tensor power spectrum, $P_t(k)$, 
to the primordial scalar 
power spectrum, $P_s(k)$, taken at wavenumber, $k_0$:

\begin{equation}
	r = \frac{P_t(k_0)}{P_s(k_0)} = \frac{A_t}{A_s}
\end{equation}
where $k_0 = 0.05 \text{ Mpc$^{-1}$}$ (see \cite{Smith2006}). For this paper we only consider the tensor-to-scalar ratio, $r$. All other 
parameters are assumed to be at their WMAP3 values (\cite{Spergel2007WMAP}). The only other parameter 
which might affect the results is the tensor spectral index, $n_t$, however we assume, in this paper, that $n_t$ is very close to zero 
(\cite{peiris03}). The problem of $n_t < 0$ will be considered in another paper.

The plan of this paper is the following. In Section \ref{wienfilt}, we 
describe the method for detection of PGWs based on measurements of the TE power spectrum. In Section \ref{simul}, we describe the numerical Monte Carlo 
simulations we tested. In Sections \ref{montecarlo}, \ref{signintro}, and \ref{wilcoxonintro}, we introduce the statistical tests used to contstrain 
PGWs. In Section \ref{comptests}, we 
compare the three different statistical tests used in this analysis. Section \ref{resul} gives 
the results for the two toy experiments described in \cite{pmkpaper1}. The only uncertainty in the first toy experiment is due to 
cosmic variance (\ref{toy1res}). In the second toy experiment, 
along with cosmic variance, we take into account instrumental noise (\ref{toy2res}). We present results of Monte Carlo simulations for 
WMAP (\ref{wmapres}) and Planck (\ref{planckres}). In Section \ref{bbcomp}, we compare signal-to-noise ratio 
for BB and TE measurements.

\section{Wiener Filtering of the TE Cross Correlation Power Spectrum} \label{wienfilt}

Wiener filtering has been used often in the case of CMB data analysis. For example, it was used to combine multi-frequency data in order 
to remove foregrounds and extract the CMB signal from the observed 
data (\cite{Tegmark1996, Bouchet1999}). Here we examine the use of the Wiener filter to subtract the PGW signal 
from the total TE correlation signal. This is done because the Wiener filter reduces the contribution of noise in a total signal by 
comparison with an estimation 
of the desired noiseless signal (\cite{vaseghibook}). In our case, the signal is the one due to PGWs only, and the signal contributed 
by density perturbations is considered to be ``noise''.

The observed signal can be written as

\begin{equation}
	C_{\ell}^{TE} = C_{\ell,s}^{TE} + C_{\ell,t}^{TE} = \left \langle a_{E,\ell m}^* a_{T,\ell m} \right \rangle
\end{equation}
where $s$ and $t$ refer to the contributions to the power spectrum due to scalar and tensor perturbations respectively. The values 
$a_{E,\ell m}$ and $a_{T,\ell m}$ refer to the 
spherical harmonic coefficients of the temperature and polarization maps. In our application to TE correlation, we consider the Wiener 
filter, $W_{TE,\ell}$: 

\begin{equation}
	W_{TE,\ell} = \frac{C_{\ell,t}^{TE}}{C_{\ell}^{TE}} = - \frac{ \left| C_{\ell,t}^{TE} \right |}{C_{\ell}^{TE}}
\end{equation}
The filtered signal, $a_{X,\ell m}^{\prime}$ (for $X = T$ and $E$), is obtained from the measured signal, $a_{X,\ell m}$, as

\begin{equation}
	a^{\prime}_{X, \ell m} =  a_{X, \ell m} W_{TE, \ell}^{1/2}
\end{equation}
In this paper, we assume the Wiener filter is perfect, in the sense that it leaves the signal due to PGWs only. We then get, for the 
filtered multipoles $C_{\ell,filt}^{TE}$,

\begin{eqnarray}
	\qquad \qquad C_{\ell,filt}^{TE} &=& \left \langle a_{T,\ell m}^{\prime *} a_{E,\ell m}^{\prime} \right \rangle \nonumber \\
	&=& W_{TE, \ell} C_{\ell}^{TE} = \frac{C_{\ell,t}^{TE}}{C_{\ell}^{TE}} C_{\ell}^{TE} = C_{\ell,t}^{TE}
\end{eqnarray}

In practice this is not true, because we are trying to 
determine $C_{\ell,t}^{TE}$, which is not known in advance. Nevertheless, the assumption that the Wiener filter is perfect is good as a first 
approximation and illustrates the 
detectability of PGWs with the help of TE correlation measurements.

The filtering can reduce the measured signal to the desired signal, but, since we are trying to remove the density perturbations and not the 
actual noise, we can not reduce the measurement uncertainties. These uncertainties in $C_{\ell}^{TE}$ are then entirely determined by 
the noise in the original signal.

From \cite{pmkpaper1}, we showed that the TE power spectrum due to PGWs is negative on large scales, hence a test determining whether 
the Wiener filtered power spectrum is negative or not is a probe of PGWs.

There are three different statistical tests we use to see if we can measure a negative TE power spectrum. The first test is a Monte Carlo 
simulation to determine signal-to-noise 
ratio, $S/N$ (Section \ref{montecarlo}). The other two tests are standard non-parametric statistical tests: the sign test 
(Section \ref{signintro}) and the Wilcoxon rank sum test (\cite{wilcoxonrst}) (Section \ref{wilcoxonintro}).

\section{Monte Carlo Simulation} \label{simul}

For all of our tests, we calculate a random variable. If the data satisfies the hypothesis that $r=0$, we can calculate the mean and 
uncertainty in the variables. If 
we make one realization of data, the random variable is determined from its distribution. Because we are not using any real 
observational data, we must run a Monte 
Carlo simulation to reduct the risk of randomly getting a value for the variable taken from the outlying area of its 
distribution. To do this, the filtered 
multipoles, $C_{\ell,filt}^{TE}$, are randomly chosen from a gaussian distribution with mean $C_{\ell,t}^{TE}$ and standard 
deviation $\Delta C_{\ell}^{TE}$ where

\begin{eqnarray}
	\qquad (\Delta C_{\ell}^{TE})^2 &=& \frac{1}{(2\ell+1)f_{sky}} \Bigg( (C_{\ell}^{TE})^2 + \nonumber \\
	&& (C_{\ell}^{TT} + N_{\ell}^{TT})(C_{\ell}^{EE} + N_{\ell}^{EE}) \Bigg)
\end{eqnarray}
(see, for example, \cite{dodelsonbook}),
the variable $f_{sky}$ refers to the fraction of the sky covered by observations
and $N_{\ell}$ is the effective power spectrum of the instrumental noise (see \cite{dodelsonbook} for details on how $N_{\ell}$ is related to actual 
instrumental noise). The underlying power spectra are generated by CAMB\footnote{see http://camb.info on web} (\cite{Lewis2000}). 

\begin{figure}
	\begin{center}
		\includegraphics[totalheight=0.25\textheight]{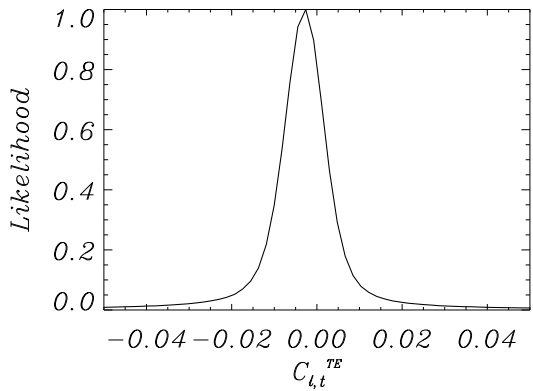}
		\\
		\caption{\small  This plot shows the combined distribution of all the $C_{\ell,t}^{TE}$. $10,000$ values for every $\ell$ in the range 
		$2 \le \ell \le 53$ were used in this histogram}
		\label{Clfig}
	\end{center}
\end{figure}

Our determination of $C_{\ell,t}^{TE}$ is dependent on $\ell$. However, for two of our tests we ignore the value of $\ell$ in the calculation of 
the random variable. We assume that the calculated random variable is gaussian. In order for this to work, the random 
variable must be calculated from gaussian variables. Fig. \ref{Clfig} shows the combined distribution of the multipoles for the ``ideal'' toy experiment 
(see Section \ref{resul}). To do this, for every $\ell$ we determine $10,000$ different 
$C_{\ell,filt}^{TE}$ values. We then 
combine all the different values, for every $\ell$, into one large distribution. Fig. \ref{Clfig} shows the errors on the multipoles are large 
enough so that for our statistical tests we can assume the multipoles are taken from a single distribution and not from a distribution 
that depends on $\ell$.

\section{Monte Carlo S/N Test} \label{montecarlo}

For this test, the random variable we calculate, $S/N$, is defined as 

\begin{equation}
	S/N = \sum_{\ell=2}^{53} \frac{C_{\ell,t}^{TE}}{\Delta C_{\ell}^{TE}}.
\end{equation}
The reason why the sum in this equation is taken in the range $2<\ell<53$ is because only in this range $sgn(C_{\ell}^{TE}(\text{scalar})) 
= -sgn(C_{\ell}^{TE}(\text{tensor}))$. In other words, if we include higher multipoles we confront with a danger of a 
false detection, because the total TE power spectrum is negative for $\ell > 53$. 

The value of $S/N$ is gaussian distributed because it is a sum of many modes of squares of 
gaussian distributed values, $C_{\ell} = a_{\ell,m}^2$. We approximate each $C_{\ell}^{TE}$ as being gaussian distributed for the 
purpose of this paper. For each set of parameters we run this simulation one million times to 
determine the mean, $\left \langle S/N \right \rangle$, and 
standard deviation, $\sigma_{S/N}$. The mean of this distribution is determined by the preassumed value of $r$, while the standard deviation 
is determined by parameters of the experiment and gives the confidence level of detection. We run such Monte Carlo simulations for 
different values of $r$ to determine in what range 
of $r$ we can detect PGWs. When then using real observational data, we can 
compare the actual value of $S/N$ with the results of Monte Carlo simulations to infer the likelihood, as function of $r$,
which determines the probability that $r \not= 0$, or that PGWs exist at detectable levels.

\section{Sign Test} \label{signintro}

The sign test is a test of compatability of observational data with the hypothesis that $r=0$. If we do have $r=0$, 
then $C_{\ell,filt}^{TE}$ will be equally distributed around zero. Application of this test to the filtered data is very simple. 
In practice, all observational data are distributed 
between several bins and the averaging of the signal is produced in each bin separately. Let $N_{bins}$ be the number of such bins. 
The sign test actually gives the 
probability that in $N_-$ bins the average is negative and 
in $N_+=N_{bins}-N_-$ it is positive, if $r=0$. 
This probability, $P$, is given by the binomial distribution

\begin{equation}
	P(N_+) = \left( \begin{array}{c} N_{bins} \\ N_+ \end{array} \right) 0.5^{N_{bins}} = \frac{N_{bins}!}{N_+ ! N_- !} 0.5^{N_{bins}}
\end{equation}

The probability that the hypothesis $r=0$ is wrong is

\begin{equation}
	P(r\not=0) \approx 1 - 2\sum_{i=0}^{N_+} P(i)
\end{equation}
The value $\sum_{i=0}^{N_+} P(i)$ is the probability that we would get $\le N_+$ positive values given $r=0$. This is the same as the 
probability of getting $\le N_+$ negative values given $r=0$. Therefore our confidence that $r\not=0$ is just $100\%$ minus the 
sum of the probabilities describe above (the probability that the $N_+$ is closer to the mean, $N_{bins} / 2$, if $r=0$). This equation 
only makes sense if $N_+ < N_{bins}/2$, since that is required for $r > 0$. 
If $N_+ > N_{bins}/2$, that would imply $r < 0$, 
which is not physical. We would have to interpret the result as a random realization of $r \ge 0$, with the most likely result of $r = 0$. Therefore 
we would not be able to say $r/not=0$ with any confidence.

Let us consider the following example: we put all measurements of $C_{\ell}^{TE}$ into $11$ bins and in three of them the average is 
positive. In this example, the 
probability that the hypothesis $r=0$ is wrong is equal to $89\%$.
 
One possible drawback of this method is that it does not take into account any measure of the signal-to-noise ratio of individual measurements. As we 
show in Section \ref{comptests}, it is possible to have 
two completely different sets of data with the same probability of having $r=0$. This test is also unable to make any prediction as to the value of $r$, 
only that it differs from zero.

\section{Wilcoxon Rank Sum Test} \label{wilcoxonintro}

This statistical test deals with two sets of data. The first set of data is taken from a real experiment which measures $C_{\ell}^{TE}$ with 
some unknown $r$. The second set of data is generated by 
Monte Carlo simulations (see Section \ref{montecarlo}) with $r=0$. The objective of the Wilcoxon rank sum test is to give the probability 
that the hypothesis $r=0$ is wrong (\cite{wilcoxonrst}). 

First, we choose some random variable $U$, whose probability distribution is known if $r=0$. For that, let us combine all data from first 
set with $n_1$ multipoles and second set with 
$n_2$ multipoles into one large data set, which obviously contains $n_1+n_2$ multipoles. Then, we rank all multipoles in the large data 
set from $1$ to $n_1+n_2$ 
according to their amplitude (rank $1$ for the smallest and rank $n_1+n_2$ for the largest). Now, the variables $R_1$ and $R_2$ are 
defined as the sum of the ranks for the first 
original data set and the second original data set, correspondingly. 
Finally, the variable $U$, is

\begin{eqnarray}
	\qquad \qquad \qquad U &=& \min(U_1,U_2) \text{, where} \nonumber \\
	U_i &=& R_i - n_i ( n_i + 1)/2 \text{, \qquad $i=1$,$2$}
\end{eqnarray}
If all multipoles of the first data set are larger than all multipoles of the second data set, then $U_1 = n_1 n_2$ and $U_2=0$. It is 
not difficult to show that $U_1 + U_2 = n_1 n_2$. 
If both sets of measurements have no evidence for PGWs, $\left \langle U_1 \right \rangle = \left \langle U_2 \right \rangle$. It is also simple
to see that $U_1 + U_2 = n_1 n_2$.

It is important to emphasize that the ranks of multipoles are random variables because 
all multipoles themselves are random variables, hence $U_1$, $U_2$, and $U$ are random variables. If $n_1 + n_2$ is large, the distribution 
of $U$ can be approximated as a gaussian 
with a known mean and standard deviation. In this approximation we have 

\begin{eqnarray}
	\qquad \qquad \qquad m_U &=& n_1 n_2 / 2 \\
	\sigma_U &=& \sqrt{\frac{n_1 n_2 (n_1 + n_2 + 1)}{12}}
\end{eqnarray}
In some cases, instead of $U$, the variables $R_1$ or $R_2$ are used. The reason $U$ is used here is because $m_U$ is symmetric 
in the data sets. If $r=0$ in both sets 
of data, then the distributions of $U_1$ and $U_2$ are the same, no matter what $n_1$ and $n_2$ are. The 
distributions of $R_1$ and $R_2$ would be the same only if $n_1 = n_2$. The probability that the first data set corresponds to $r \not= 0$ 
obtained from the test in which 
$R_1$ or $R_2$ is used is the same as if $U$ is used. 

Since this test requires Monte Carlo simulations for the second set of data, we ran this test many times for many different data sets 
to get an accurate mean value for $U$. 

To reject the hypothesis $r=0$ means to detect PGWs. Using the Wilcoxon rank sum test the allowable value of $r$ is determined, if
instead of comparing with simulated data with $r=0$, we compare with simulated data with $r=r_0 \not= 0$. 
In order to get a range of allowable values for $r$, we need to run multiple Monte 
Carlo simulations with multiple values for 
$r_0$. This is where the assumption that the $C_{\ell,t}^{TE}$ are from a random distribution that is independent of $\ell$ is 
used (see Section \ref{simul}). This implies that the ranks are random 
variables. If the errors on the $C_{\ell,t}^{TE}$ are small enough, then the ranks will be predetermined. Therefore, our assumption 
about the distribution of $U$ will not be true and 
the test would have to be modified. Fortunately, this is not the case for even an experiment only limited by cosmic variance.

To illustrate how this test works, let us consider the following example. Assume there are $4$ multipoles in the first set of data and 
consider that $r=0.3$ is the correct value. There are also 
$4$ multipoles in the second set of data (which for sure corresponds to $r=0$). All quantities below are expressed in $\mu$K$^2$. 
The value for the first data set are
$C_{10}^{TE} = -0.005$, $C_{20}^{TE} = 0.02$, $C_{30}^{TE} = -0.015$, and $C_{40}^{TE} = -0.01$. The values for the second data set are 
$C_{10}^{TE} = 0.03$, $C_{20}^{TE} = 0.003$, $C_{30}^{TE} = -0.02$, and $C_{40}^{TE} = -0.003$. A ranking of multipoles gives the ordering from lowest 
to highest, with $1$ referring to the first data set and $2$ referring to the second data set, as $21112212$. This results in 
$R_1 = 2 + 3 + 4 + 7 = 16$, $U_1 = 16 - 10 = 6$, and 
$U_2 = 16 - 6 = 10$. Therefore $U=\min\{10,6\}=6$. For $n_1 = n_2 = 4$, to reject the hypothesis that $r=0$ at $95\%$ confidence 
level, $U_1$ should be less than one (see, for example, \cite{lehmannbook}). 
In this example, since $U_1 = 6 > 1$, the first set of data cannot be considered as a detection of PGWs.

\section{Comparison of Tests} \label{comptests}

The $S/N$ test is greatly affected by outlying measurements. A measurement of one large negative multipole could falsely implay a 
detection. Both the sign test and the 
Wilcoxon rank sum test are not affected by individual outlying measurements. In the sign test, the value of individual measurements 
is irrelevant, because the test is 
sensitive only to the sign of individual measurements. The Wilcoxon rank sum test is affected by outliers, but considerably less than 
the $S/N$ test. If the outlier 
is larger (or smaller) than every other multipole, its rank does not depend on its particular value.

If we have two completely different sets of data, the main disadvantage of the sign test, as mentioned in Section \ref{signintro}, is 
that it could give the same result, while for the two other tests the 
chance to obtain the same value of $r$ is negligible. For example, one set of data, consisting of $4$ small negative multipoles and 
$4$ large positive multipoles, gives 
the same result as another set of data, consisting 
of $4$ large negative multipoles and $4$ small positive multipoles. The $S/N$ test gives two very different values of $S/N$ for these 
two sets of data. We can also use the 
Wilcoxon rank sum test to compare these two sets of data. In this case $U = 16 = \frac{1}{2}m_U$, which corresponds to a confidence 
level of hypothesis that $r=0$ of less than $10\%$.

With observational data, the sign test can be applied and does not require any Monte Carlo simulations (which 
could be considered as an advantage of this test). The $S/N$ test requires Monte Carlo simulations, but only for the 
distribution of the random variable $S/N$. The Wilcoxon rank sum test requires large Monte Carlo simulations and combines the data sets 
generated by these simulations with observational data. In other words, Monte Carlo simulations are absolutely necessary after 
obtaining observational data, which may be considered a disadvantage of this test.
Thus, each of the three tests has advantages and disadvantages, suggesting that the best way to work out observational 
data is to apply all these three tests.

\section{Discussion and Results} \label{resul}

The current best limit of $r$ is $r<0.3$ at $95\%$ confidence provided by 
WMAP in combination with previous experiments (\cite{Spergel2007WMAP}). We need to see 
if this method can detect a value of $r$ that is currently within the limit. We provide results for $r=0.3$ in order to see 
how well these tests can detect an 
amount of PGWs that is very close to being ruled out by BB measurements. 
For all our tests we assumed that there is no foreground contamination. For the experiments that are not 
observing the full sky, correlations between multipoles must be taken into account. We bin together the highly correlated multipoles 
so that the correlations between the bins are 
sufficiently small.

The two toy experiments that were used in \cite{pmkpaper1} are also used here to constrain $r$. 
The two toy experiments are fully described in \cite{pmkpaper1}, however we will also reproduce their description here. 

The first toy experiment is a full sky experiement. For this experiment, we take measurements 
over the full sky with no instrumental noise. The only uncertainty will be due to cosmic variance. This experiment represents the 
best limit to which the gravitational waves can be detected with the TE correlation. A space-based experiment with access to 
the full sky is the closest to this experiment. It is similar to what the Beyond Einstein inflation probe will be able to detect. 
This toy experiment will be hereafter referred to as the ideal experiment. 

The second toy experiment is a more realistic experiment.  Measurements of the CMB are taken on $3\%$ of the sky in one 
frequency ($100$ GHz) for $3$ years. The 
noise in each detector of the $50$ polarization sensitive bolometer pairs can be described by their noise equivalent temperature (NET) of 
$450$ $\mu K \sqrt{s}$. The detector beam 
profiles are assumed to be gaussian and and it is described by their full width half maximum (FWHM) of $0.85^{\circ}$.

This second toy experiment is similar to current ground-based experiments. Constraints from this experiment represent those 
that can and will be obtained in the next several years using this method. This will be referred to as the realistic experiment.

We also look at experiments with instrumental noise similar to the satellite experiments WMAP and Planck. 
The predicted errors for Planck are based on using the $100$ GHz, $143$ GHz, and $217$ GHz channel in the High Frequency Instrument (HFI). The numbers are 
gotten from the Planck science case, the ``bluebook''\footnote{http://www.rssd.esa.int/index.php?project=Planck}. The WMAP noise was obtained by 
using $3$ years of the Q-band, V-band, and W-band detectors. 


\subsection{Ideal Experiment} \label{toy1res}

\begin{figure}
	\begin{center}
		\includegraphics[totalheight=0.25\textheight]{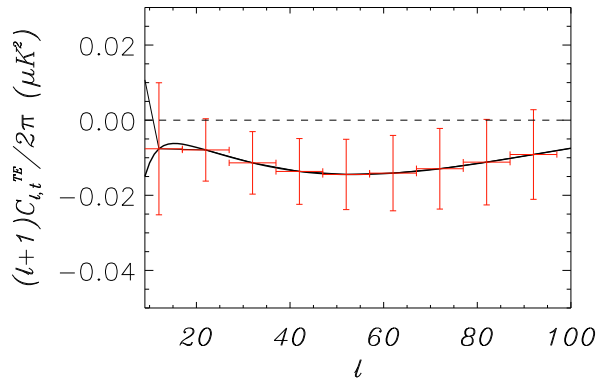}
		\\
		\caption{\small  The black line is the TE mode due to PGWs with $r=0.3$. Red is the error bars for the ideal experiment 
		calculated from the total TE power spectrum 
		binned in intervals of $\Delta \ell = 10$.}
		\label{exp1r3}
	\end{center}
\end{figure}

A plot of the TE power spectrum due to PGWs for $r=0.3$ and $n_t=0.0$ for the ideal experiment is shown 
in Fig. \ref{exp1r3}. The errors bars in Fig. \ref{exp1r3} are calculated from the total TE power spectrum. There are no correlations 
between multipoles, because we observe the 
full sky, but we plot the error bars binned in intervals of $\Delta \ell = 10$ for simplicity in the plot.

\begin{figure}
	\begin{center}
    	\includegraphics[totalheight=0.25\textheight]{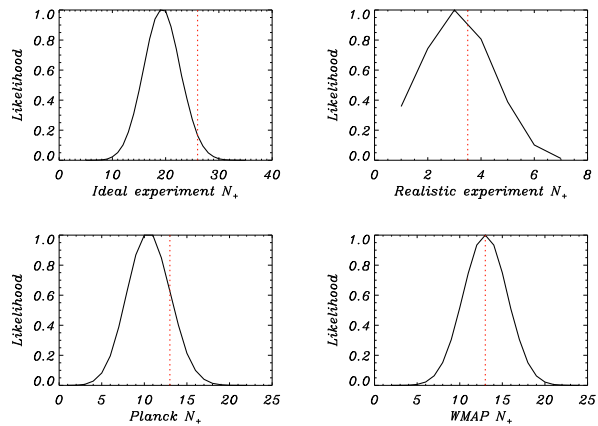}
    	\\
    	\caption{\small This is a plot of the distribution of the number of positive multipoles for the Monte Carlo simulation for 
		the ideal experiment (upper left), 
		the realistic experiment (upper right), Planck (lower left), and WMAP (lower right). The dotted red line shows where $N_+ = \frac{1}{2}N_{bins}$}
		\label{signdist}
    \end{center}
\end{figure}

The Monte Carlo simulation gives an average of $19$ measured TE power spectrum multipoles greater than zero out of a 
total of $52$ independent multipoles. If the null hypothesis was true, 
the sign test would indicate there is a $3.5\%$ chance of measuring $\le 19$ positive multipoles. This is equivalent to a
$\approx 1.8\sigma$ detection. 
A plot of the distribution of the number of positive multipoles is shown in the upper panel plot of Fig. \ref{signdist}. There is an $81\%$ chance for the 
observed $N_+$ to give a $1\sigma$ detection of PGWs. 

\begin{figure}
	\begin{center}
    	\includegraphics[totalheight=0.25\textheight]{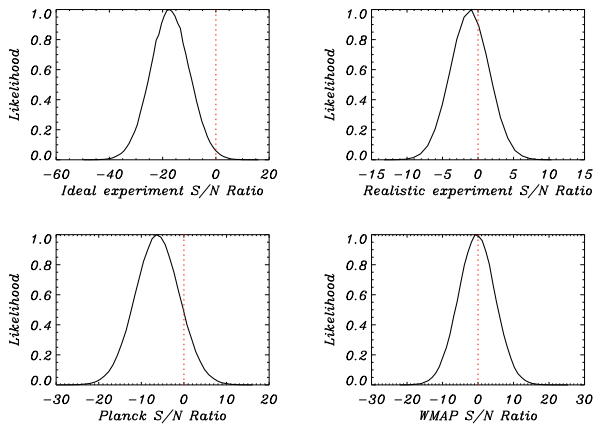}
    	\\
    	\caption{\small The $S/N$ statistic distribution for the ideal experiment (upper left), realistic experiment (upper right), 
		Planck (lower left), and WMAP (lower right). The dotted red line shows where $S/N=0$.}
		\label{SNPlot}
    \end{center}
\end{figure}

\begin{figure}
	\begin{center}
    	\includegraphics[totalheight=0.25\textheight]{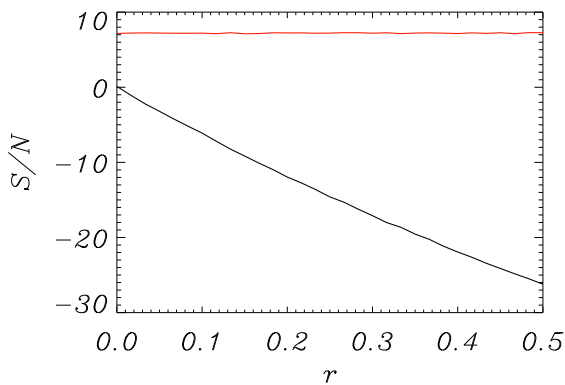}
    	\\
    	\caption{\small This is a plot of $\left \langle S/N \right \rangle$ and $\sigma_{S/N}$ as a function of $r$ for the ideal 
		experiment. The black line is $\langle S/N \rangle$ 
		and the red line is $\sigma_{S/N}$.}
		\label{SNmeanSD}
    \end{center}
\end{figure}

The $S/N$ test gives a mean value of $S/N = -17.1$ and standard deviation of 
$7.21$. The upper left panel in Fig. \ref{SNPlot} shows the 
distribution of the $S/N$ values for the Monte Carlo simulation with $r=0.3$. If $r=0.3$ we would have a $0.8\%$ probability of 
the measured $S/N > 0$. This negative value signifies that a non-zero tensor-to-scalar ratio produced an anti-correlation. We can assume 
that the standard deviation would be the same if the mean of $S/N$ was $0$ (equivalent 
to $r=0.0$), because it is equivalent to adding a constant value to every measured value (and hence adding a constant to $S/N$ which 
would not change the error). Therefore, if $r=0$, the 
probability of getting $S/N < -17.4$ is $0.8\%$, and hence we have a $99\%$ chance that $r \not= 0$. A plot of 
$\left \langle S/N \right \rangle$ and $\sigma_{S/N}$ 
as a function of $r$ is shown in Fig. \ref{SNmeanSD}. As can be seen from the plot, we can predict a value of $r$ for any value of 
$S/N$. The value of $\sigma_{S/N}$ is a 
relatively constant function of $r$ and so our prediction about the distribution of $S/N$ for different value of $r$ is a good 
approximation to the true distribution.

\begin{figure}
	\begin{center}
    	\includegraphics[totalheight=0.25\textheight]{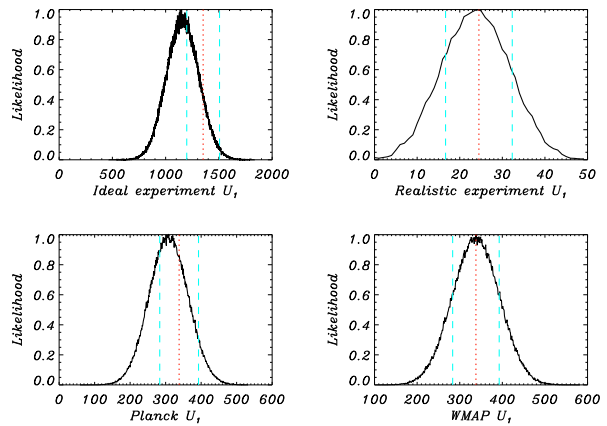}
    	\\
    	\caption{\small This is the plot of the distribution of $U$ for the ideal experiment (upper left), realistic experiment 
		(upper right), Planck (lower left), and 
		WMAP (lower right). The red dotted line is the value for $m_U$ and the light blue dashed lines enclose the $1\sigma$ 
		region for $U$ assuming the hypothesis that $r=0$}
		\label{WilcoxonPlot}
    \end{center}
\end{figure}

The Wilcoxon rank sum test gives $U_{avg}-m_U = -1.23\sigma_U$. The variable $U_{avg}$ is the 
mean value for $U$ in the Monte Carlo simulations described earlier. The 
values $m_U$ and $\sigma_U$ are given in Section \ref{wilcoxonintro}. The distribution of $U$ for the Monte Carlo simulations with 
$r=0.3$ is shown in Fig. \ref{WilcoxonPlot}. The standard 
deviation of the distribution of measured $U$ is the same as the standard deviation of the distribution of $U$ assuming the hypothesis 
that $r=0$. The only difference between 
the distributions is that $m_U$ is shifted by a constant value. Therefore, there is a $22\%$ chance 
that $U-m_U < -2\sigma_U$. There is also a $40\%$ chance that we measure $U-m_U < -1\sigma_U$, and are not even able to make a 
$1\sigma$ detection of PGWs.

\begin{figure}
	\begin{center}
    	\includegraphics[totalheight=0.25\textheight]{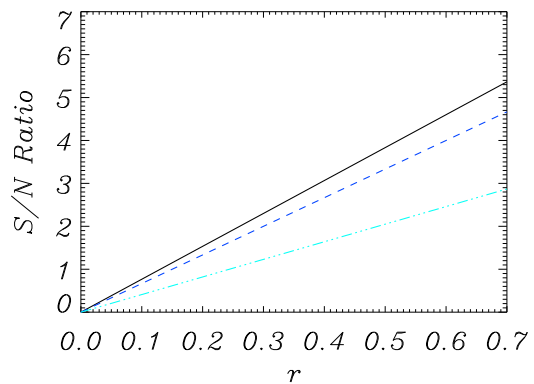}
    	\\
    	\caption{\small This is the plot of the signal-to-noise ratio (number of $\sigma$s) for different values of $r$ for the 
		three different tests. The black line is the $S/N$ test, the dashed dark blue 
		line is the sign test, and the dotted-dashed light blue line is the Wilcoxon rank sum test}
		\label{Detectability}
    \end{center}
\end{figure}

A comparison of the three tests is shown in Fig. \ref{Detectability}. This is obtained by simulated with with several values of $r$ 
and then interpolating between them. A $2\sigma$ detection is 
obtained for $r=0.26$ ($S/N$ test), $r=0.3$ (sign test), and $r=0.5$ (Wilcoxon rank sum test), highlighting its intended use as 
a monitor of a false positive detection for large $r$.

\subsection{Realistic Ground Based Experiment} \label{toy2res}

\begin{figure}
    \begin{center}
    	\includegraphics[totalheight=0.25\textheight]{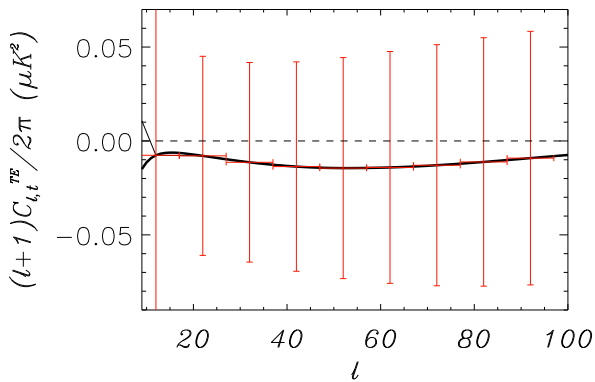}
    	\\
		\caption{\small This figure is the same as Fig. \ref{exp1r3} except for the realistic experiment}
		\label{exp2r9}
	\end{center}
\end{figure}

A plot of the TE power spectrum due to PGWs with $r=0.3$ is shown in Fig. \ref{exp2r9}. The error bars are calculated from the 
total TE power spectrum. Observations of an incomplete 
sky require the multipoles to be binned in sizes of $\Delta \ell = 10$. This experiment has much 
larger error bars than the ideal experiment and it will not be able to detect low values of $r$ with the 
TE cross correlation only. 

The results for the Wiener filtering method were much worse than those for the ideal experiment for $r=0.3$. Since this experiment 
observes a small portion of the 
sky, the multipoles are correlated and we must bin together to get reasonably uncorrelated measurements. For this experiment, we 
only have $7$ to $8$ uncorrelated multipoles, 
instead of $52$ uncorrelated multipoles in the case where the full sky is observed. Getting $7$ out of $8$ negative multipoles is 
a $3\%$ probability if there are no 
PGWs. For the Monte Carlo simulations of the realistic experiment, on average, half of measured multipoles are positive and half 
are negative. A plot of the distribution of the number of positive multipoles is shown in the upper right panel of Fig. \ref{signdist}. In this 
case, we cannot distinguish $r=0.3$ from $r=0.0$ with any significance. 

The $S/N$ test gives an average value of $S/N=-0.95$ with 
standard deviation of $2.64$. For the realistic toy experiment, the distribution of $S/N$ for $r=0.3$ is shown in the upper right panel of Fig. 
\ref{SNPlot}. In order to obtain $68\%$ confidence detection of 
PGWs, we must use $r \approx 0.7$. In this sense the TE test provides monitoring and insurance against false positive detection with $r > 0.7$, 
which could arise, for example, if foregrounds or other systematic effects arer improperly removed. 

The last statistical test, the Wilcoxon rank sum test, gives $U_{avg} - m_U = -0.20 \sigma_U$. The distribution of $U$ for $r=0.3$ is 
shown in the upper right panel of 
Fig. \ref{WilcoxonPlot}. This gives the weakest result in terms of the three tests for the Wiener filtered data. 
The realistic experiment will not be able to constrain $r < 0.3$ using the TE cross correlation power spectrum. Its 
limit is closer to $r < 0.7-0.9$ at only $68\%$ confidence depending on the test used. For a higher confidence in a detection of PGWs, 
the value of $r$ would need to be much higher. Since the observed distribution of 
$U$ corresponds almost exactly to the simulated distribution of $U$ under the assumption that $r=0$, therefore we have a $16\%$ chance of 
measuring $U-m_U < -1\sigma_U$. 

\subsection{WMAP} \label{wmapres}

The results of the Wiener filtering showed that the WMAP cannot make a detection 
of gravitational waves using the TE cross correlation power spectrum alone. As with the two toy experiments, the result of the 
scalar and tensor separation was similar.
The Monte Carlo simulation gave on average gave $13$ positive multipoles out of a total of $26$ uncorrelated multipoles. We 
would get the same result if the input data had 
$r=0.0$ so we cannot detect PGWs with WMAP using only the TE power spectrum. A plot of the distribution of the number of 
positive multipoles is shown in the lower right panel of 
Fig. \ref{signdist}. As can be seen, this distribution of $N_+$ for WMAP noise and $r=0.3$ is simply the distribution for $r=0$.  

For WMAP, the S/N test gives the value of $S/N=-0.02$ with a standard deviation of 
$5.09$. The distribution is shown in the lower right panel of Fig. \ref{SNPlot}. The distribution is centered around $S/N=0$ 
so there is no chance of using this 
test to detect PGWs in WMAP's TE power spectrum. The probability of getting a $1\sigma$ or $2\sigma$ detection is the same 
probability that we would randomly get a detection if 
there are no PGWs.

The rank sum test gives a value of $U_{avg}-m_U = -0.004\sigma_U$, 
which is implies no ability to distinguish WMAP's observed TE data from a data set with no PGWs. 
A plot of the distribution of $U$ for WMAP error bars is shown in the lower right panel of Fig. \ref{WilcoxonPlot}. We reach the  
same conclusion for WMAP noise as for 
the realistic experiment. There is only a $16\%$ chance that we can measure $U-m_U < -1\sigma_U$ and make a $1\sigma$ detection of $r=0.3$

The published WMAP results show an anti-correlation of TE power spectrum at large scales. Unfortunately this is not a detection of PGWs as 
theorized in \cite{baskaran06}. The contribution to the TE power spectrum due to PGWs only changes sign once for $\ell \lesssim 90$. If a claimed 
evidence for 
gravitational waves is to be believed, then the TE power spectrum would have to change sign three times for $\ell \lesssim 60$. In fact, other than 
the two anticorrelations at low $\ell$, the rest of 
the multipoles, up to $\ell=53$, are consistent with $r=0$. None of the described tests applied to the current WMAP data will give any detection of PGWs.

\subsection{Planck} \label{planckres}

The sign test gives on 
average $10$ positive measurements of the TE power spectrum out of a total of $26$ uncorrelated multipoles. There is a $16\%$ chance 
of getting $\le 10$ positive 
multipoles if $r=0$. A plot of the distribution of the number of positive multipoles for Planck is shown in the lower left panel of 
Fig. \ref{signdist}. There is a $50\%$ 
chance that we will measure $N_+ < 10$ and hence have a $1\sigma$ detection of $r=0.3$.

The $S/N$ test gives a value of $S/N=-6.24$ with 
a standard deviation of $5.09$. There is only a $10\%$ chance that the $S/N$ test results in a value of $S/N$ larger than zero, 
if $r=0.3$, and a $10\%$ chance getting $S/N < -3.12$ if $r=0$. 
This is close to a $90\%$ probability of detection. The distribution of the $S/N$ variable is shown in lower left panel of Fig. \ref{SNPlot}. 

Again, the rank sum test gives the lowest confidence result with a value of $U_{avg}-m_U = -0.66\sigma_U$. A plot of the distribution 
of $U$ is shown in the lower 
left panel of Fig. \ref{WilcoxonPlot}. There is a $37\%$ probability that we will measure $U-m_U < -1\sigma_U$ and a $9\%$ probability that we 
measure $U-m_U < -2\sigma_U$ for Planck.

\section{Comparison of Measurements of TE Power Spectrum with BB Power Spectrum} \label{bbcomp}

CMB polarization can be separated into two distinct components: E-mode (grad) polarization and 
B-mode (curl) polarization. PGWs generate B-mode polarization in contrast to density perturbations (see for example 
\cite{Seljak97Pol, selzal97, Kamionkowski98}). Therefore, 
most CMB polarization experiments searching for evidence of PGWs focus on measuring the 
BB power spectrum (see \cite{CloverIntro, BowdenOpt, BicepPaper}). For different aspects of CMB polarization generated by PGWs see, for example, 
\cite{basko, Polnarev85, Crittenden93, Frewin94, colefrewinpolnarev95, kamionkowski97, 
Seljak97Pol, selzal97, Kamionkowski98, baskaran06, Keating2006}. 

We have shown that the TE cross correlation power spectrum offers another method of detecting PGWs (\cite{crittenden94}). 
The TE power spectrum is two orders of magnitude 
larger than the BB power spectrum and it was originally suggested that it might be easier to detect gravitational waves in the TE power 
spectrum than using the BB power spectrum (\cite{baskaran06, grishchuk07}). However, as we have shown in 
\cite{pmkpaper1}, that, when we use zero multipole method, uncertainties in TE measurements exceed uncertainties in BB measurements in such a 
way that the signal-to-noise ratio for BB measurements is 
better than for TE measurements. We show below that the same is true for the separation of scalars and tensors.

However, an advantage of TE measurements for ground based experiments, which observe only a small fraction of the sky, is related to 
the fact that the main spurious effects in the BB power spectrum are 
cuased by E/B mixing. This will limit the $r$ that 
can be detected (\cite{challinor05}). The E-modes are practically unaffected by E/B mixing, so, in contrast to the BB measurements, the TE power spectrum 
should be nearly the same for both full and partial sky measurements.

Along with this, the methods based on the TE cross correlation can be considered as very useful 
auxiliary measurements of PGWs, because
systematic effects in TE measurements are independent from those in BB measurements. For example, T/B leakage or even E/B leakage could swamp a
detection of BB, whereas T/E leakage would be small and well
controlled (see \cite{meir07}). These BB systematics could falsely imply a large $r$, but measurements of the TE power spectrum
provide insurance against such a spurious detection. Additionally, galactic foreground contamination will
affect BB and TE in different ways, which enable us to perform powerful cross-checking and subtraction of foregrounds in BB measurements.

For the zero multipole method, we explained why the signal-to-noise ratio for BB measurements is 
better than for TE measurements. Below we give simple summarizing arguments why the same is true for the Wiener filtering of 
the TE power spectrum. 

If $N_{\ell} \ll C_{\ell}^{BB}$, the signal-to-noise ratio for the BB power spectrum is

\begin{equation}
	(S/N)_{BB} =  \frac{C_{\ell}^{BB}}{\Delta C_{\ell}^{BB}} = \gamma \frac{C_{\ell}^{BB}}{C_{\ell}^{BB} + N_{\ell}} \approx \gamma,
\end{equation}
where

\begin{equation}
	\gamma =  \sqrt{\frac{(2\ell+1)f_{sky}}{2}}
\end{equation}
If $N_{\ell} > C_{\ell}^{BB}$ then we will not be 
able to detect PGWs and a comparison with the TE power spectrum is not worthwhile. 

If $N_{\ell} \ll C_{\ell}^{EE}$ and $r < 1$, for the TE power spectrum, the signal-to-noise ratio is

\begin{eqnarray}
	(S/N)_{TE} &=& \frac{C_{\ell,t}^{TE}}{\Delta C_{\ell}^{TE}} \nonumber \\
	&=& \sqrt{2} \gamma \frac{C_{\ell,t}^{TE}}{\left[ (C_{\ell}^{TE})^2 + (C_{\ell}^{TT} + N_{\ell}/2)(C_{\ell}^{EE} + N_{\ell})\right]^{1/2}} \nonumber \\
	&\approx& \sqrt{2} \gamma  \frac{C_{\ell,t}^{TE}}{\left[ (C_{\ell}^{TE})^2 + C_{\ell}^{TT}C_{\ell}^{EE}\right]^{1/2}} \nonumber \\
	&\approx& \sqrt{2} \gamma \frac{r}{\alpha + \beta r}
\end{eqnarray}
where $\alpha$ and $\beta$ are

\begin{eqnarray}
	\alpha &=& \frac{\sqrt{(C_{\ell,s}^{TE})^2 + C_{\ell,s}^{TT} C_{\ell,s}^{EE}}}{D_{\ell}^{TE}}, \nonumber \\
	\beta &=&  \frac{2C_{\ell,s}^{TE} D_{\ell}^{TE} + D_{\ell}^{TT} C_{\ell,s}^{EE} + C_{\ell,s}^{TT} D_{\ell}^{EE}}{2 D_{\ell}^{TE} \alpha}, 
\end{eqnarray}
where
\begin{equation}
	D_{\ell}^{XY} = C_{\ell,t}^{XY} / r
\end{equation}
One can see that $\alpha$ and $\beta$ are on the order of unity. Therefore, the signal-to-noise ratio is approximated as

\begin{equation}
	(S/N)_{TE} = \sqrt{2} \gamma \frac{r}{\alpha+\beta r} \approx \sqrt{2} \gamma \frac{r}{\alpha}
\end{equation} 
In other words if $r < \alpha/\beta \sim 1$, BB measurements have the obvious advantage in comparison with the Wiener filtering of 
the TE power spectrum. Indeed if $r \lesssim 1$, 
$(S/N)_{BB} \sim \gamma$, while $(S/N)_{TE} \sim \gamma r < \gamma$. This is because in BB measurements, applying proper data 
analysis, we can entirely eliminate contributions of scalar perturbations to CMB polarization signal as well as to the uncertainties. 
For the perfect Wiener filtering of the TE power 
spectrum, we can eliminate the contribution of scalar perturbations to the signal only, but cannot eliminate their contribution to the uncertainties.

\section{Conclusion}

The method described here is one in which we filter out the signal due to density perturbations, leaving only the contribution to the 
TE power spectrum due to PGWs. We 
then test the resulting TE power spectrum to see if it is negative. 
Three different statistical tests were used to see if there was a significant detection of PGWs. The $S/N$ test can 
give a value for $r$ using a comparison with Monte Carlo simulations, while the Wilcoxon rank sum test can only give an allowable range for $r$. The 
sign test will only tell us if $r \not= 0$. 

From \cite{pmkpaper1}, we saw that we could detect $r=0.3$ to $3\sigma$ with a measure of $\ell_0$, the position where the TE power spectrum 
first changes sign, for the ideal 
experiment. Using the method discussed in this paper, we see that we are unable to make this significant of a detection. The best 
result was for the $S/N$ test which would 
give a $2.3\sigma$ detection of $r=0.3$. To detect PGWs on the level of $3\sigma$, the tensor-to-scalae ratio $r$ should be $r \ge 0.4$. The 
sign test would give $2\sigma$ detection for $r=0.3$ and a 
$3\sigma$ detection for $r=0.45$. The Wilcoxon ranked sum test gives only a $1.2\sigma$ 
detection for $r=0.3$ and a $3\sigma$ detection for $r=0.7$. Similar results were gotten for the other three experiments tested. Thus 
in the sense of potential to detect PGWs, the zero 
multipole method is the best, next best is the $S/N$ test, then the sign 
test, and the worst is the Wilcoxon ranked sum test. 

\cite{baskaran06} present illustrative examples in which high $r$ is consistent with measured TT, EE, and TE correlations. The 
value of $r$ is so high in these examples  
that if PGWs with such $r$ really existed, current BB experiments would already detect PGWs. 
All models predict that the TE cross correlation power spectrum change sign only once for $\ell < 100$. The fact WMAP cannot exclude 
several multipoles with $C_{\ell}^{TE} > 0$ in between multipoles of $C_{\ell}^{TE} < 0$ means that the TE cross correlation power 
spectrum either changes sign several times for $\ell < 100$ or there is some instrumental noise which causes some anticorrelation measurements. Using 
instrumental noise consistent with WMAP, our Monte Carlo simulations give $\Delta \ell_0 \approx 16$ and $\ell_0 > 40$, which means 
that there is no evidence of 
PGWs in the TE correlation power spectrum. 

\section*{Acknowledgments}

NJM would like to thank the Astronomy Unit, School of Mathematical Science at Queen Mary, University of London for hosting 
him while working on this paper. We acknowledge helpful comments on this manuscript by Kim Greist and Manoj Kaplinghat.
BGK gratefully acknowledges support from NSF PECASE Award AST-0548262. We acknowledge using CAMB to calculate the power spectra in 
this work.

\label{lastpage}

\bibliography{Inflation,Polarization}
\bibliographystyle{mn2e}
\end{document}